2MPo2C-01

# REBCO delamination characterization by 90 degree peel test

Jun Lu, Jeremy Levitan, Aliya Hutley, and Hongyu Bai

***Abstract*— REBCO tape has successfully used in ultra-high field magnets. Mechanically, it is very strong in its length direction but is prone to delamination in the thickness direction. In an epoxy impregnated REBCO magnet, thermal strain alone could delaminate the conductor. Even for dry wound REBCO coil, a conductor with very low delamination strength is still a concern. Therefore, it is important to characterize the delamination strength of the conductor. In the past decade, significant progresses have been made in the characterization of REBCO delamination strength. Among several developed characterization methods, the 90 degree peel test is simple to setup and seems to offer reproducible results. Therefore, this test could be used as a reliable method to characterize the relative delamination strength of REBCO tapes for quality control purposes.**

**This paper presents our development of 90 degree peel test method for quality assurance test of the 40 T all-superconducting magnet project at the National High Magnetic Field Laboratory. We investigated the factors that influence the test results, such as thickness and RRR of the copper layer. We found that peel strength increases with decreasing Cu thickness. We also found that peel strength is positively correlated with RRR of the copper. Despite these effects, 90 degree peel test is still a valuable quality assurance tool to evaluate delamination strength for large volume of tapes with the same copper thickness.**

***Index Terms*—REBCO, delamination, peel strength.**

## I. INTRODUCTION

RARE Earth barium copper oxide (REBCO) coated conductor is an excellent conductor for high field superconducting magnets. REBCO coated conductor consists of multi layers films grown on a mechanically strong substrate. Therefore, it can tolerate high mechanical stress in its length direction, but it is prone to delamination in the thickness direction. The delamination is especially concern for epoxy impregnated coil where it was demonstrated that the thermal stress from the coil cool down alone can result in considerable critical current degradation due to delamination [1]. The delamination strength of REBCO tapes has been measured by transverse tensile tests (anvil test) [2], cleavage tests [3], double cantilever beam test [4] and peel tests [5], [6]. In this paper, we report a 90 degree peel test method that is developed for large volumes of quality assurance program for the 40 T all-superconducting magnet project at the National High Magnetic Field Laboratory. The factors influence the test results are discussed.

This paragraph of the first footnote will contain the date on which you submitted your paper for review, which is populated by IEEE. This work was performed at the National High Magnetic Field Laboratory, which is supported by National Science Foundation Cooperative Agreement No. DMR-1644779, DMR-1839796, DMR- 2131790, and the State of Florida. *Corresponding author: Jun Lu*.

Jun Lu, Jeremy Levitan, Aliya Hutley and Hongyu Bai are with the National High Magnetic Field Laboratory, Tallahassee, FL 32310 USA.

## II. EXPERIMENT

The samples of about 100 mm long were taken from SuperPower SCS4050-AP tapes, which is 4 mm wide with 50 μm Hastelloy substrate and electroplated copper 20 μm per side. The surrounding Cu at the edges were trimmed by using a pair of scissors. After manually pre-peeling from one end for about 15 mm, the substrate side of the conductor was attached to a holder made of G-10 bar by adhesive tapes as shown in Fig. 1. The G-10 bar was placed horizontally with the pre-peeled Cu layer of the REBCO facing down. A small variable weight (small beaker) was hung from the pre-peeled 15 mm Cu. the weight was gradually increased at a rate about 0.5 g/s by dripping water to the beaker until the peeling started. The peel strength is defined as the load (weight) needed to start peeling divided by the tape width. The test method was calibrated using Scotch® Magic™ tape with a known peel strength. Our measured peel strength for the Magic™ tape is 1.06 N/cm, consistent with the reported peel strength of the tape of 1.1 N/cm [7].

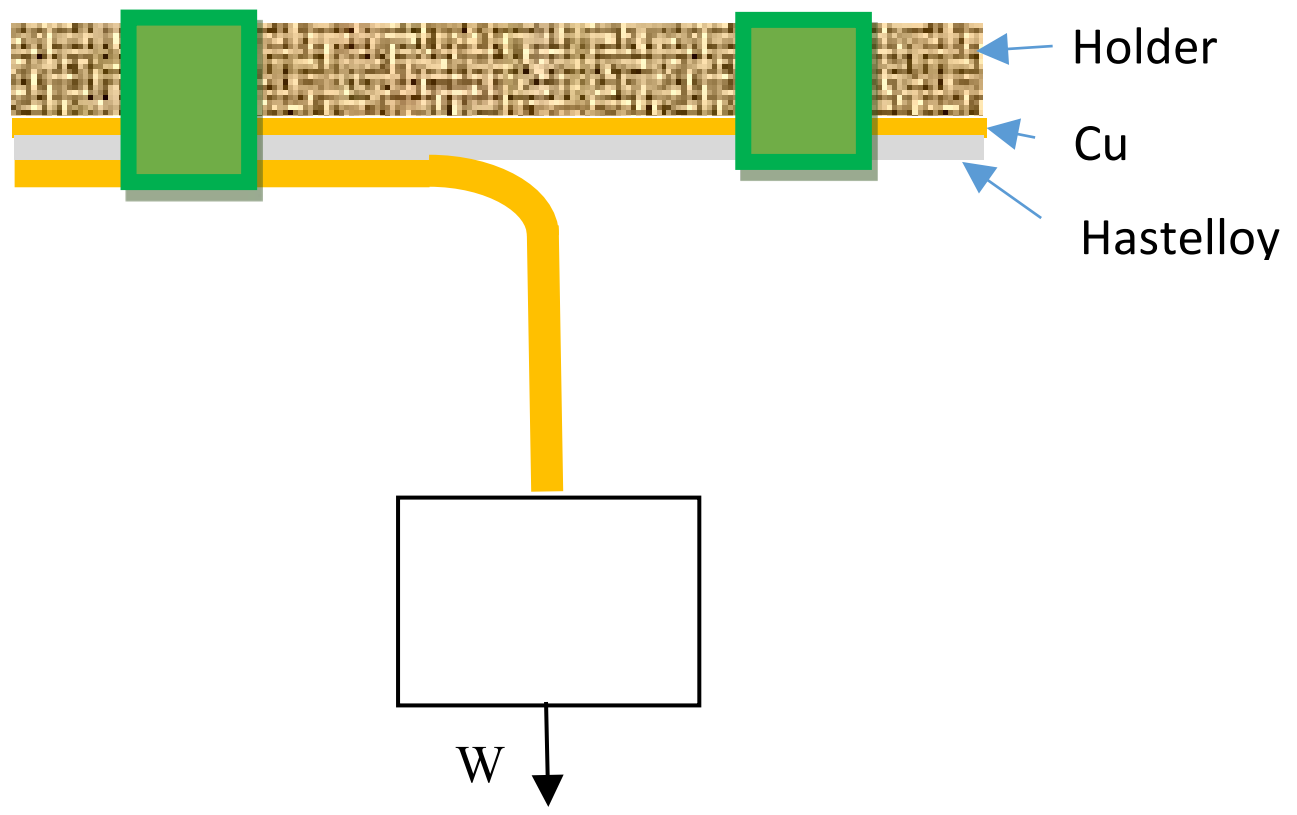

**Fig. 1.** A schematic of the 90 degree peel test device.

It should be noted that this test method is not the same as ASTM – D6862 [8] which measures the peeling force at a given peeling speed. Whereas the peeling force measured by this method is the minimum force needed to start the peeling motion. For relative evaluation of delamination strength between samples, our test method should be adequate.

## III. RESULTS

### *A. Repeatability*

To evaluate the repeatability of peel test, several samples were cut adjacent to each other from one piece length. The results are shown in Fig. 2(a). The average peel strength is 0.85 N/cm with a standard deviation of 0.09 N/cm. The variation is reasonably small. Samples were also taken from the front end and back end of each of several spools of conductor each 100 - 200 meters long. The differences between front end and back end, as shown in Fig. 2(b), are moderate.

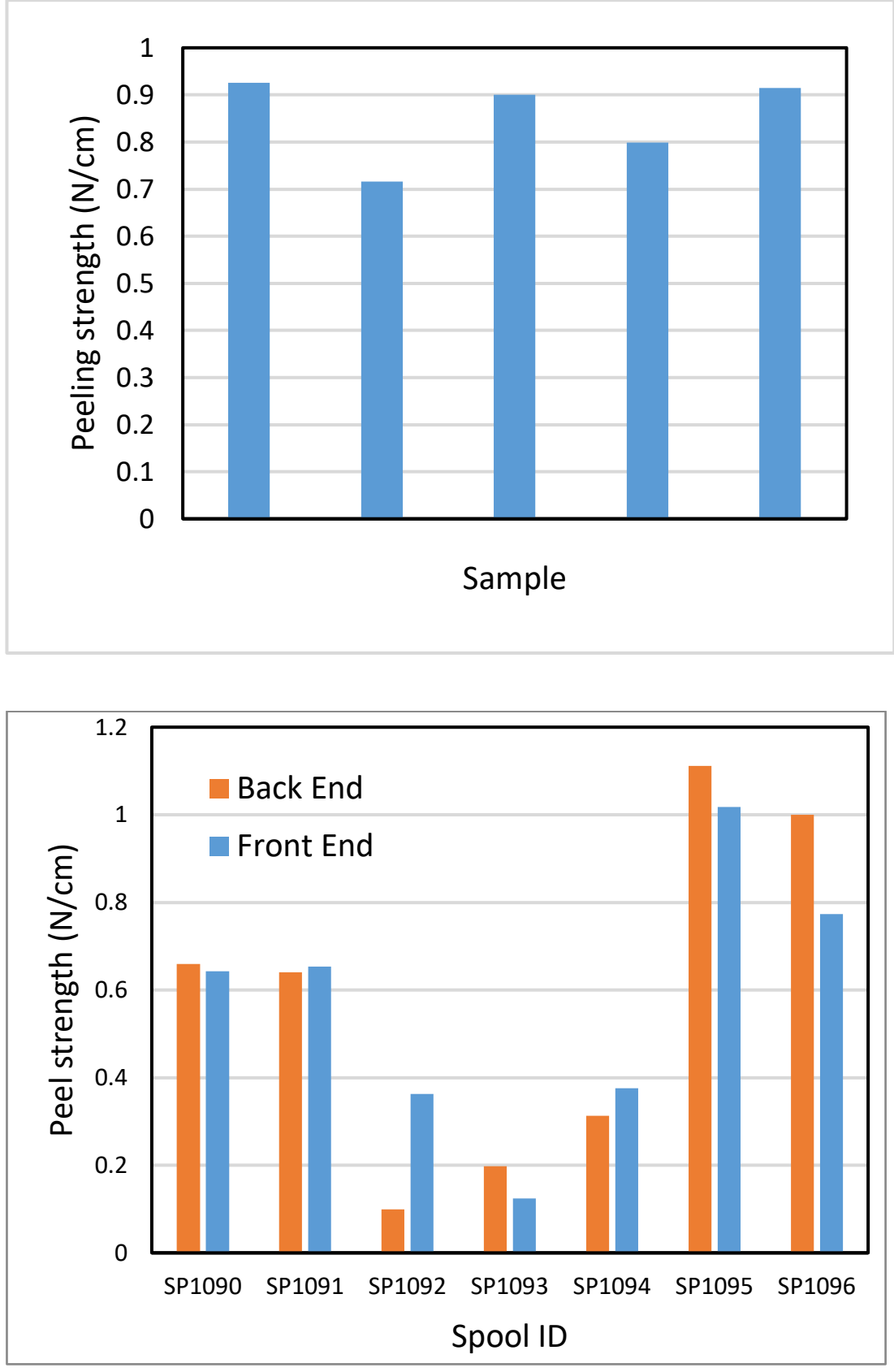

**Fig. 2.** (a) Repeatability of 5 samples cut from one piece length, (b) Front to end variations for several spools, each 100 – 200 m long.

### *B. Surface after peel test*

In most of cases, peeled surfaces on both sides were black color indicating residual REBCO on both sides of the delamination as shown in Fig. 3(a). However, for samples with very low delamination strength, substrate side surface has yellowish color (Fig. 3(b)) which is due to the interference of light from interfaces of the buffer layers which are transparent. Our transmission electron microscopy (TEM) study confirmed these observations [9].

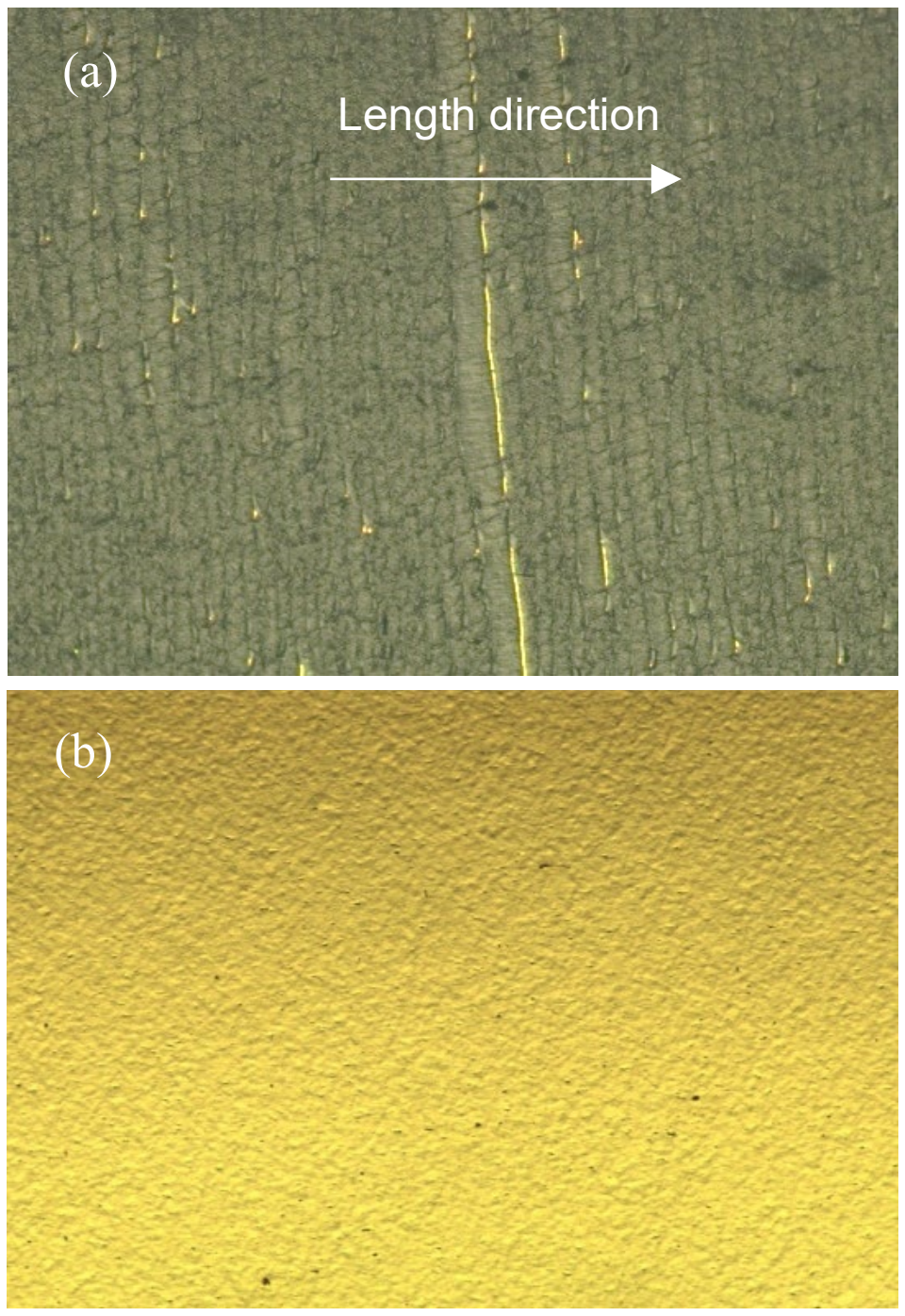

**Fig. 3.** Surface of the substrate side after peel test. (a) Most cases with residual REBCO. (b) Surface of one of buffer layer in the case of very weak peel strength,

### *C. Effect of copper thickness*

The influence of film thickness on peel test has been well documented [10]-[15]. We studied the effect of copper thickness by preparing multiple samples from the same piece length and removing various thicknesses of copper (starting at 20 µm) by using copper etchant APS-100. The results are plotted in Fig. 4.

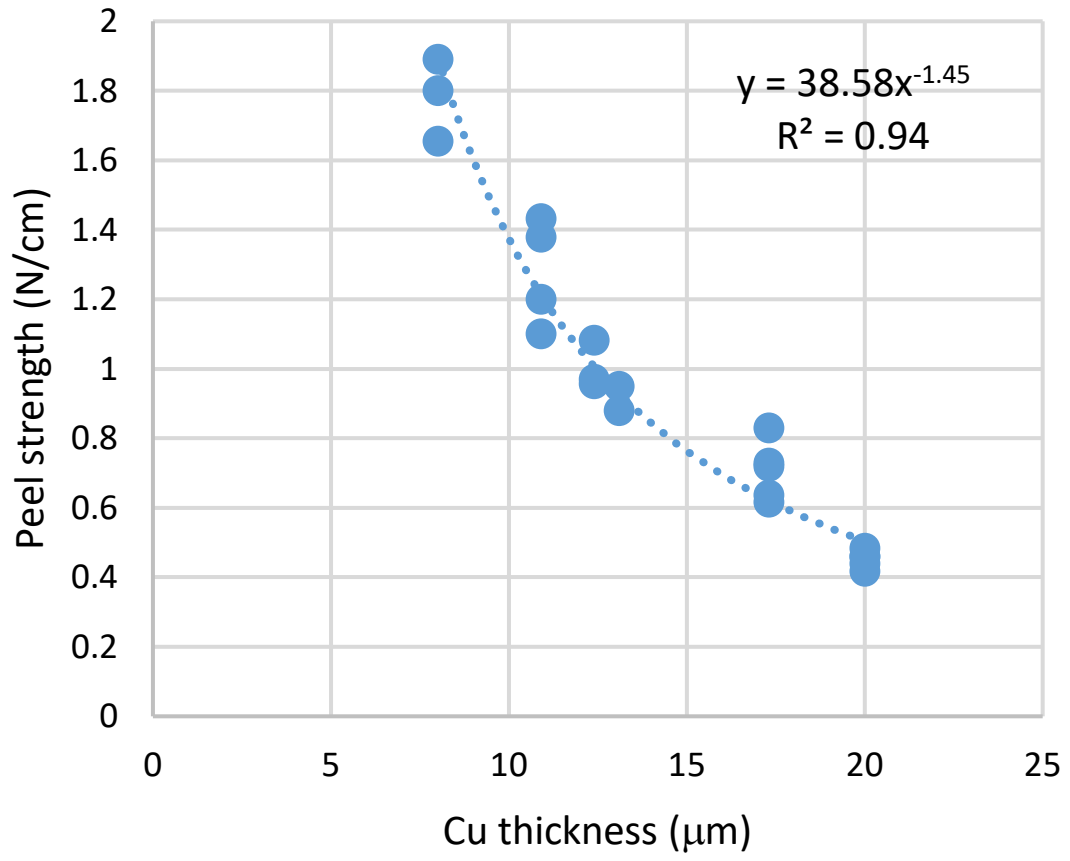

**Fig. 4.** Effect of Cu thickness on peel strength. The data are fitted with a power function (dash line).

Evidently the peel strength increases with decreasing Cu thickness consistent with previously reported peel test of Cu-Cr film on silicon [10], [11]. Since the interfaces are nearly identical for these samples, the different peel strength strongly suggests that the measured peel strength is not sole property of the interfaces. It contains a significant contribution from the bending of the copper layer. Hence the value of peel strength cannot be used directly to represent delamination property of REBCO layers. However, for quality assurance of large volume of REBCO tapes such as in the case of the 40 T project where Cu thickness (20 μm) is the same for all samples, the relative value of peel strength can be used to identify the sample with relatively low delamination strength.

### *D. Correlation with Cu RRR*

We studied the correlation between the peel strength and other quality assurance data of REBCO, and found surprisingly that peel strength is correlated with RRR of the copper. Fig. 5(a) shows peel strength and RRR data of 74 spools of conductor (total of over 10 km). High peel strength appears to correspond to high RRR except for a few cases. Fig. 5(b) shows the correlation between the two.

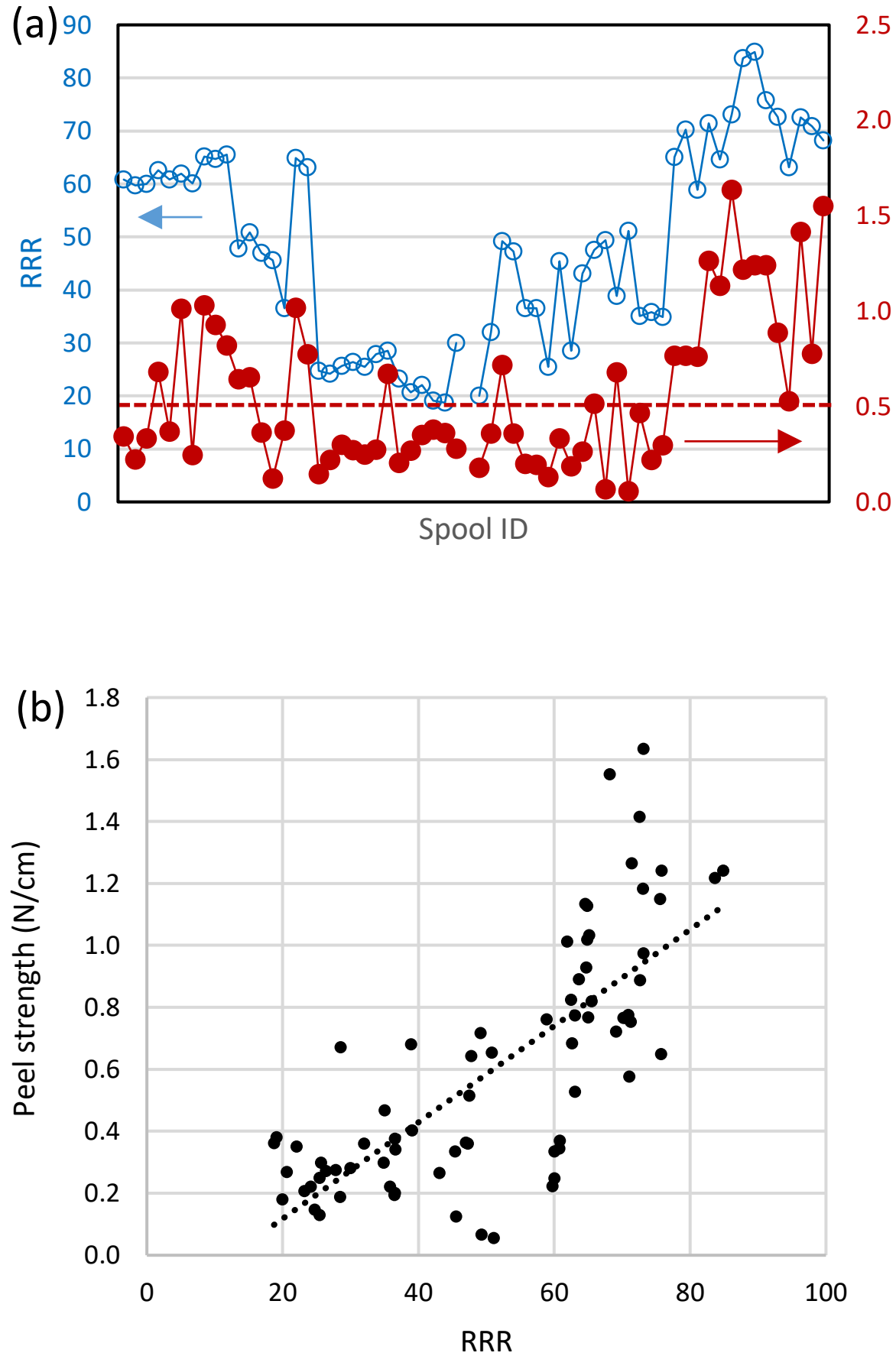

**Fig. 5.** Correlation between peel strength and Cu RRR. (a) RRR and peel strength for multiple spools of conductors. The dash line represents peel strength from [5] for comparison. (b) Peel strength vs. RRR plotted from data in (a). The dash line is the linear regressio.

To further understand the effect of the RRR, a few samples were annealed in argon for 30 minutes at different temperatures. Annealing results in Cu grain growth which leads to higher RRR due to the reduced grain boundary scattering [16]. Fig. 6 plots RRR and peel strength as functions of anneal temperature. Since the grain size growth by annealing also reduces copper's yield strength, this experiment strongly suggests that higher peel strength in Fig. 5(a) is the consequence of the softer copper layer.

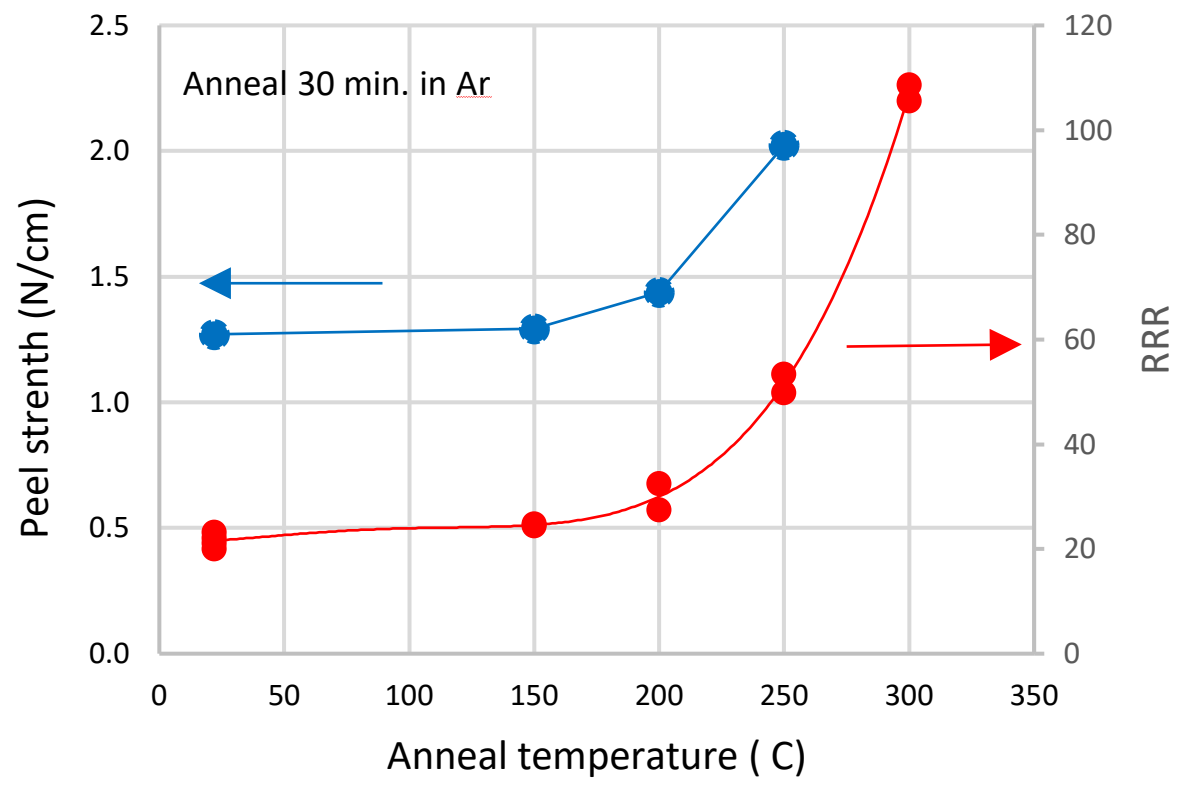

**Fig. 6.** Effect of annealing. Peel strength and Cu RRR are plotted against anneal temperature. The lines are guides to the eye.

## IV. DISCUSSIONS

Conductor delamination has been a concern for high field REBCO magnets. In addition to the possible thermal stress in an epoxy impregnated magnet, the electromagnetic force during the operation might delaminate the conductor, as recent experimental study showed delamination by electromagnetic force alone [17]. Therefore, a test that can identify conductor with weak delamination is very important.

The peel test developed in this work has the advantage of applicable to quality control and quality assurance process to quickly identify conductors with low delamination strength. However, the measured peel strength value is not an intrinsic material property. It is significantly influenced by thickness and mechanical property of the copper layer. Therefore, the test is useful only for relative comparison of conductor with the same Cu thickness and RRR.

### *A. Cu thickness and stiffness dependence*

The detailed mechanics of the peel test is very complex [15]. While polymer films showed higher peel strength with increasing film thickness [12]-[14], metal films [10], [11] shows the opposite trend which is consistent with our results.

The effect of Cu RRR is unexpected. Based on the results of our annealing experiment, it might be explained as the following. Higher RRR is indicative of larger grain size which is linked to lower the yield strength. This results in lower bending stiffness. It is the low bending stiffness that leads to higher peel strength. The effect of the copper property was previously noticed in Ref. [5] where the copper plated by different systems resulted in different peel strengths. The effect of the bending stiffness is also demonstrated by testing peel strength with the pre-peeled Cu side being attached to the G-10 bar and weight being applied on the substrate side. In such case, the measured peel strength was considerably lower than the regular tests where the substrate side was attached to the G-10 bar. Evidently with the same interfacial structure stiffer bending arm (the substrate) results in lower peel strength.

Figure 7 is a schematic depicting two cases where the dash line indicates the location of the crack tip. The soft and thin Cu takes less load to bend. However, the distance *d* between the crack tip and the load (the downward arrow) is shorter. The load needed to create the same critical stress intensity factor at the crack tip is therefore larger. Overall, it requires larger load to propagate the crack.

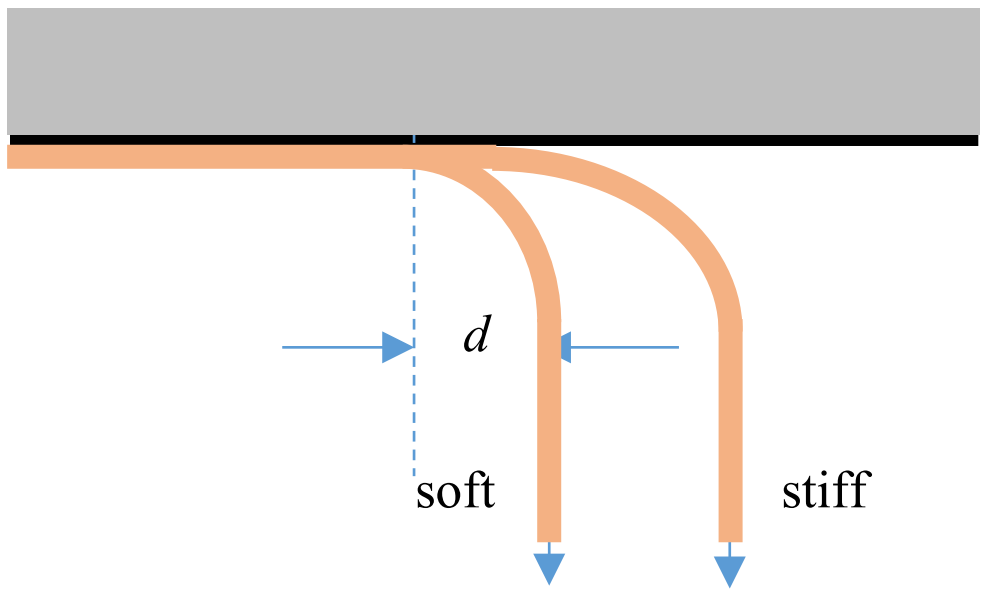

**Fig. 7.** peel strength in cases of different copper stiffness.

Evidence suggests that the measured peel strength *P* (in N/m or J/m$^2$) consists of a term of interfacial energy *G* (J/m$^2$) and a term of plastic bending energy *Q* [10],

$$P = G + Q \quad (1)$$

Here *Q* is a function of Cu thickness and RRR. It should be noted that *Q* is interrelated to *G* and *P*. For example, when interfacial energy *G* = 0, both peel strength *P* and bending energy *Q* would be zero.

In practice, if the Cu is thinner than 10 µm, it would be difficult to perform peel test which is limited by Cu strength in the longitudinal direction. On the other hand, for thick Cu (for example > 50 µm for SuperPower tapes), the thickness effect renders peel strength too low to be a useful indicator of interfacial delamination strength.

### *B. Normalized peel strength*

It is possible to normalize the peel strength considering the effects of both Cu thickness and RRR. In our cases, since the copper thickness is specified as 20 µm, the peel strength is normalized to RRR = 50 by multiplying a factor of RRR/50. Here we assume peel strength is approximately proportional to RRR which is supported by the data shown in Fig. 5(b).

The normalized peel strength is plotted in Fig. 8. For the 74 samples shown in the figure, the average of normalized peel strength is 0.57 N/cm with a standard deviation of 0.24 N/cm. The normalized peel strength allows us to identify a few spools of conductors with very low peel strength, as labelled by the circles in the figure.

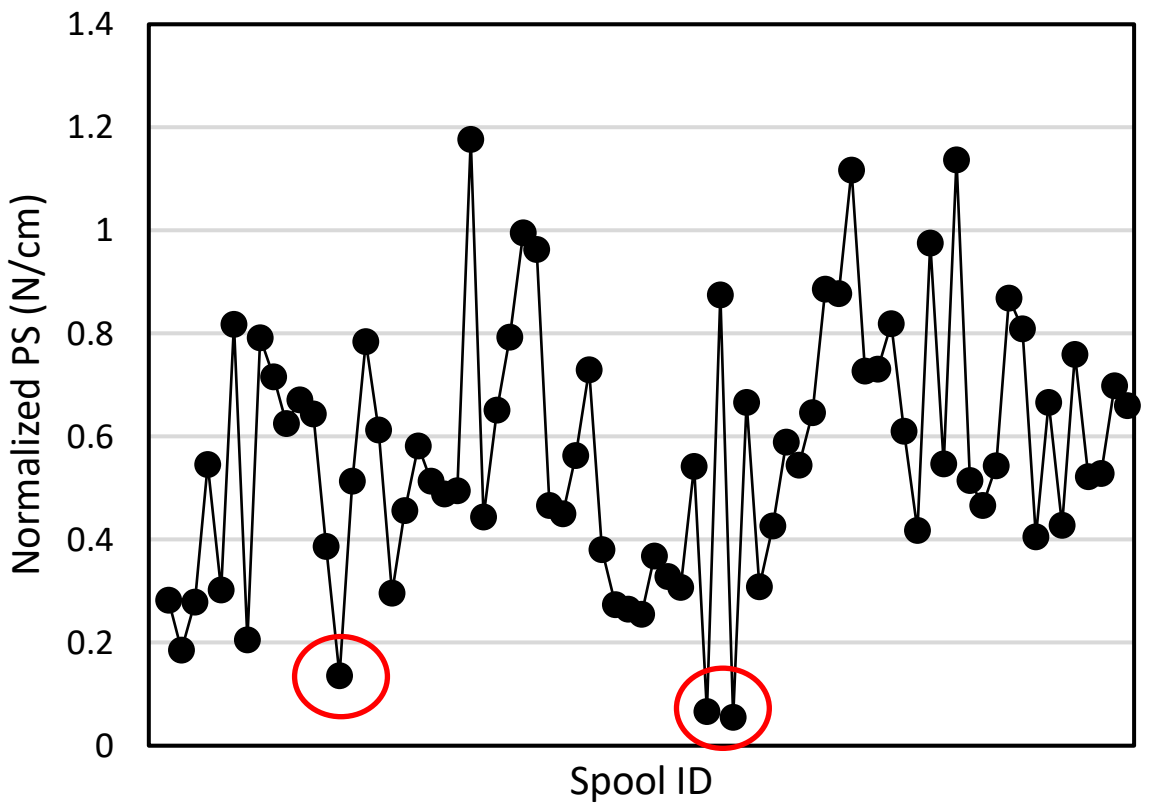

**Fig. 8.** Peel strength data in Fig. 5(a) are normalized to RRR = 50. A few spools of very low peel strength are identified (circled).

### *C. Correlation with other delamination tests*

The 90 degree peel test can be conveniently used to identify weak delamination strength within large volume of REBCO conductors with the same thickness of copper made by the same process. In principle, this can also be achieved by other delamination test methods [2]-[6]. Therefore, it is highly desirable to establish a positive correlation between multiple delamination test methods by testing the same set of strong and week delamination samples with these methods. The preliminary experiments towards this direction have been discussed [18]. The other important question is, how the measured delamination strengths can be used to establish a magnet design criterion. To answer this question, a comparative study with epoxy impregnated coils seems to be necessary.

## V. Conclusion

The 90 degree peel test is developed to characterize delamination strength of REBCO conductor for quality assurance of the 40 T all-superconducting magnet project at the National High Magnetic Field Laboratory. The measurements were carried out for SuperPower REBCO tapes with a 20 µm Cu stabilizer. The measurement is repeatable. It is found, however, that the peel strength is strongly correlated with thickness and RRR of the Cu layer. The reasons for the correlation are discussed. The measured peel strength is normalized to RRR = 50. A few conductors with very low peel strength are identified. It is demonstrated that peel test is an effective quality assurance tool to weed out REBCO tapes with very weak delamination strength.

## References

[1] T. Takematsu, *et al.*, “Degradation of the performance of a YBCO-coated conductor double pancake coil due to epoxy impregnation,” *Phys. C, Supercond. Appl.*, vol. 470, no. 17, pp. 674–677, 2010.

[2] Hyung-Seop Shin “Evaluation of delamination characteristics in GdBCO CC tapes under transverse load using anvil test methods for various anvil contact configurations at 77 K” *Supercond. Sci. Technol.* vol. 32, 2019, Art. no. 104001.

[3] Y. Yanagisawa, et al., "Remarkable weakness against cleavage stress for YBCO-coated conductors and its effect on the YBCO coil performance", *Physica C,* vol. 471, pp. 480–485, 2011.
[4] T. Miyazato, et al., 'Mode I type delamination fracture toughness of YBCO coated conductor with additional Cu layer', *Physica C*, vol. 471 pp. 1071-1074, 2011.
[5] Y. Zhang, et al., 'Adhesion strength study of IBAD-MOCVD-based 2G HTS wire using a peel test', *Phys. C, Supercond*. vol. 473, pp. 41–47, 2012.
[6] N Long et al, "Mode I Delamination Testing of REBCO Coated Conductors via Climbing Drum Peel Test", *IEEE Trans. Appl. Supercond.*, vol. 28, no. 4, June 2018, Art. no. 6600705.
[7] D. Yu, et al., "Tailored polyurethane acrylate blend for large-scale and high-performance micropatterned dry adhesives", *J. Mater. Sci.,* vol. 54, 2019, Art. no. 12925.
[8] *Standard Test Method for 90 Degree Peel Resistance of Adhesives,* ASTM D6862 – 11, ASTM international, 2021.
[9] Y. Xin, et al., '1J-ML-Or1A-05: Atomic-scale observation of cracks in REBCO coated conductor due to delamination', in Applied Superconductivity Conference, Salt Lake City, UT, Sept. 1-6, 2024.
[10] J. Kim, et al., 'Mechanical effect in peel adhesion test', *J. Adhesion Sci Technol*, vol. 3, no. 3, pp. 175-187, 1989.
[11] K.S. Kim and J. Kim, 'Elasto-plastic analysis of the peel test for thin film adhesion', *Transactions of ASME*, vol. 110, pp. 266, 1988.
[12] M. Ciaverella, et al., 'A general expression for the maximum force in peeling a tape from a rigid substrate with an initial crack', *J. Adhesion*, vol. 99, no. 6, pp. 1031-1043, 2023.
[13] H.B. Yin, et al., 'Determination of the interface properties in an elastic film/substrate system', *Int. J. Solids and Structures*, vol. 191–192, pp. 473–485, 2020.
[14] Z. Peng, et al., 'Peeling behavior of a viscoelastic thin-film on a rigid substrate', *Int. J. Solids and Structures,* vol. 51, pp. 4596–4603, 2014.
[15] M.D. Bartlett, et al., 'Peel Tests for Quantifying Adhesion and Toughness: A Review', *Progress in Materials Science*, pp. 101086, 2023.
[16] J. Lu, et al., 'Characterization of residual-resistance-ratio of Cu stabilizer in commercial REBCO tapes' *Cryogenics*, vol. 141, 2024, Art. no. 103901.
[17] M. Bonura, et al, '2OrSM-4: Quench concomitant to Lorentz-force-induced delamination in commercial REBCO coated conductors', in MT-28, Aix-en- Provence, France, Sept. 10-15, 2023.
[18] H.S. Shin, private communications.